\documentclass[twocolumn,prl,aps,showpacs]{revtex4}

\usepackage{graphicx}
\def\be{\begin{equation}}
\def\ee{\end{equation}}
\def\bea{\begin{eqnarray}}
\def\eea{\end{eqnarray}}

\begin{document}

\title{Zero-energy vortex bound states in noncentrosymmetric superconductors} 

\author{Chi-Ken~Lu and Sungkit~Yip}

\affiliation{Institute of Physics, Academia Sinica, Nankang,
Taipei 115, Taiwan}

\date{\today }

\begin{abstract}

We consider bound states at the vortex core of a non-centrosymmetric
superconductor.  We show that, despite the mixing of singlet and
triplet order parameters, zero energy states survive within certain
parameter space as in vortices of some chiral p-wave states.

\end{abstract}


\maketitle

Noncentrosymmetric superconductors, such as CePt$_3$Si,
Li$_2$Pt$_3$B, and Mg$_{10}$Ir$_{19}$B$_{16}$, have drawn a great
deal of interest in the past few
years\cite{Noncentro_Ex1,Noncentro_Ex2,Noncentro_Ex3}.
Microscopically, the degeneracy among the two pseudospin states at a
given momentum, connected by the composite operations of
time-reversal and inversion transformations, is lifted since the
latter symmetry is broken. Hence,  Cooper pairs can no longer be
classified in terms of singlets versus triplets, nor $s$-wave versus
$p$-wave \cite{Gorkov}. In CePt$_3$Si, for example, the
superconducting state is believed to have a triplet order parameter
${\bf{d}(k)}\propto(k_x\hat{\rm{y}}-k_y\hat{\rm{x}})$ in addition to
a singlet one \cite{Noncentro_Ex1}. It has been proposed that this
mixing is  responsible for some of the seemingly contradictory
behaviors of these superconductors \cite{Noncentro_Ex2,HayashiNMR}.
Moreover, some intriguing broken symmetry properties of these
superconductors have been predicted \cite{microYip}.

Recently, bound state \cite{deGennes} within the odd winding number
vortex of the p-wave $k_x+ik_y$ superfluid has been a hot topic
since among them there is a localized zero-energy
state\cite{Kopnin,Ivanov,Tewari,Volovik}, which, in terms of its
associated creation operator, is a self-hermitian Majorana fermion
\cite{RG}. Proposals that utilize such zero-energy states for
implementing the topological quantum computation\cite{Ivanov,Tewari}
are based on its unique properties including robustness against
perturbations from deformations of order parameters and nonmagnetic
impurities\cite{Volovik,RG,Stone,Radzihovsky,IndexTheorem}.

It is well known that the bound state spectrum associated with an
isolated vortex with winding number one in an $s$-wave
superconductor is $E=(n+\frac{1}{2})\omega$  with the energy scale
$\omega$ of order $\Delta^2/E_F$\cite{deGennes}  (here $n$ is an
integer, $\Delta$ the energy gap and $E_F$ the Fermi energy), while
the spectrum in the A phase of $p$-wave superfluid $^3$He is given
by $E=n\omega$ which includes a state at zero energy \cite{Kopnin}.
In this paper, we shall study the vortex bound states in
noncentrosymmetric superconductors described by a mixture of singlet
and triplet order parameters, which is intermediate between the
previous two examples. First, we demonstrate that there are two such
zero-energy states (for each $p_z$, to be defined later)
corresponding to the pure $p$-wave order parameter with
${\bf{d}(k)}\propto(k_x\hat{\rm{y}}-k_y\hat{\rm{x}})$ presenting the
combined up and down equal-spin pairings. Thus the two states differ
by spin orientation, and at first sight, would be coupled by the
additional $s$-wave order parameter and hence acquire finite
energies. However, we shall show that, as long as the $s$-wave
component is smaller than a critical value, the zero-energy states
survive. Besides, we note that this condition is identical to the
existence of a nodal gap. In addition, we consider a Rashba
spin-orbital interaction.  We shall show that the zero-energy states
again survive.

The creation operator $\alpha^{\dag}$ for a quasiparticle
excitations in an inhomogeneous superconductor is a linear
combination of electronic annihilation and creation field operators
$\psi_{\uparrow,\downarrow}(\vec r)$ and
$\psi_{\uparrow,\downarrow}^{\dagger}(\vec r)$:

\be
    \alpha^{\dag}=\int_{\vec{r}}\left(u_{\uparrow}(\vec{r}),u_{\downarrow}(\vec{r})\right)
    \left(\begin{array}{cccc}\psi^{\dag}_{\uparrow}(\vec{r})\\\psi^{\dag}_{\downarrow}(\vec{r})
    \end{array}\right)+
    \left(v_{\uparrow}(\vec{r}),v_{\downarrow}(\vec{r})\right)
    \left(\begin{array}{cccc}\psi_{\uparrow}(\vec{r})
    \\\psi_{\downarrow}(\vec{r})
    \end{array}\right)\:,
\ee satisfying $[H_{\rm{eff}},\alpha^{\dag}]=\epsilon\alpha^{\dag}$
where $H_{\rm eff}$ is the effective mean-field Hamiltonian and
$\epsilon$ is the quasiparticle energy. The coefficients
$(\hat{u},\hat{v})=(u_{\uparrow},u_{\downarrow},v_{\uparrow},v_{\downarrow})$
satisfy the Bogoliubov deGennes (BdG) equation,

\be
    \left(\begin{array}{cccc}&H_0&\Pi\\
    &\Pi^{\dag}&-H_0^*\end{array}\right)
    \left(\begin{array}{cccc}\hat{u}\\\hat{v}\end{array}\right)
    =\epsilon\left(\begin{array}{cccc}\hat{u}\\\hat{v}\end{array}\right)\label{BdG}
\ee where, for the ordinary cases, $ H_0 = -\frac{\nabla^2}{2m}-E_F$
is the kinetic energy (we shall add the possible Rashba interaction
later). $\Pi$ is a two-by-two matrix due to the pairing.  In the
singlet case $\Pi$ represents $\Delta_s(\vec{r})(i\sigma_2)$. In the
triplet case
$\Pi=\frac{1}{2}(\vec{\nabla}\cdot\vec{D})+\vec{D}\cdot\vec{\nabla}$.
The vector
$\vec{D}=-i\vec{\nabla}_{{\bf{k}}}\Delta(\vec{r},{\bf{k}})$ where
$\Delta(\vec{r},{\bf{k}})$ is the order parameter parameter. By the
conventional notation,
$\Delta(\vec{r},{\bf{k}})=\Delta_p(\vec{r}){\bf{d}(k)}\cdot{\vec{\sigma}(i\sigma_2})$.
$\Pi$ is then a sum
of the above two when both order parameters are present.

We first consider the simpler case ({\it c.f.} e.g. \cite{Tewari})
where the order parameter is of pure $p$-wave character with
${\bf{d}(\bf{k})}=(k_x\hat{\rm{y}}-k_y\hat{\rm{x}})/p_F$. (Here $p_F
\equiv ( 2 m E_F)^{1/2}$). We shall show that, for each value of
momentum along the vortex line $p_z$ less than $p_F$, there are two
zero-energy states with the associated wavefunction
$(\hat{u}(\vec{r}),\hat{v}(\vec{r}))^T$ given by

\bea
    (1,0,-1,0)^TR_1(\rho)e^{ip_zz}\:,\nonumber\\
    (0,e^{i\phi},0,-e^{-i\phi})^TR_2(\rho)e^{ip_zz}\:,\label{solution1}
\eea in cylindrical coordinates $\vec{r}=(\rho,\phi,z)$, where the
radial functions $R_{1,2}$ are independent, finite, and decaying at
infinity.

For an isolated vortex line with winding number 1, the order
parameter can be expressed as
$\Delta_p(\vec{r})=\Delta_p(\rho)e^{i\phi}$ where we shall choose
the gauge where $\Delta_p(\rho)$ is real and positive. The coupling
$\Pi$ in (\ref{BdG}) is then

\be
    \frac{\Delta_p(\rho)}{p_F}\left(\begin{array}{cccc}
    &e^{-i\phi/2}(\partial_{\rho}-\frac{i}{\rho}\partial_{\phi})e^{i\phi/2}&0\\
    &0&e^{i3\phi/2}(\partial_{\rho}+\frac{i}{\rho}\partial_{\phi})e^{i\phi/2}
    \end{array}\right)\:.\label{p_order}
\ee  $\Delta_p(\rho)$ is zero at $\rho=0$ and increases toward its
asymptotic value $\Delta_0$ within a range of coherence length
$\xi \gg p_F^{-1}$. In principle we should add also respectively
$\frac{1}{2 p_F}\partial_{\rho}\Delta(\rho)$
 and $\frac{1}{2 p_F} e^{2 i\phi}\partial_{\rho}\Delta(\rho)$
to the
upper left and lower right elements in eq (\ref{p_order}) due to the
$\vec \nabla \cdot \vec D$ term in $\Pi$, but it can be shown that,
since these terms are regular $\rho \to 0$ and vanish necessarily as
$\rho \to \infty$, they do not affect the arguments below and so we
would not show them explicitly to simplify the equations. It is
clear from (\ref{p_order}) that the BdG equation (\ref{BdG}) for
excitations with up and down spins become decoupled. For the states
with $\epsilon=0$, we denote the two independent excitations by
$\alpha^{\dag}_1$ and $\alpha^{\dag}_2$. For $\alpha^{\dag}_1$ with
up spin, the wavefunctions can be factored into
$u_{\uparrow}(\vec{r})=e^{ip_zz}u_{\uparrow}({\rho})$ and
$v_{\uparrow}(\vec{r})=e^{ip_zz}v_{\uparrow}({\rho})$. The equations
for $u_{\uparrow}+v_{\uparrow}$ and $u_{\uparrow}-v_{\uparrow}$
become decoupled:

\be
    \left[\frac{1}{2m}(
    \frac{d^2}{d{\rho}^2}+\frac{1}{\rho}\frac{d}{d{\rho}})
    +\tilde \mu
    \mp\frac{\Delta_p({\rho})}{p_F}(\frac{d}{d{\rho}}+\frac{1}{2{\rho}})\right]
    (u_{\uparrow}\pm{v_{\uparrow}})=0\:,\label{eq_ex1}
\ee where $\tilde \mu (p_z) \equiv E_F-\frac{p_z^2}{2m}$ is an
effective chemical potential. Here we assume $p_z<p_F$, and thus
$\tilde \mu>0$. Two properties about Eq.\ (\ref{eq_ex1}) have to be
noted here. First, for each of the above equation the presence of a
regular singular point at the origin forces the general solutions to
be divergent at the origin. Second, the equation corresponding to
$u_{\uparrow}+v_{\uparrow}$ with a minus sign in front of
$d/d{\rho}$ has its both solutions unbounded at infinity, whereas
the equation for $u_{\uparrow}-v_{\uparrow}$ has two solutions
decaying as $\rho\rightarrow\infty$. The above can be shown by
substituting $e^{ip_{\|}{\rho}}$ into the asymptotic form of
(\ref{eq_ex1}) where terms proportional to $1/\rho$ can be neglected
and $\Delta_p(\rho)$ is replaced by $\Delta_0$. Then we have
$-x^2+\frac{\mu}{E_F}\mp{i}\frac{\Delta_0}{E_F}x=0$, where $x\equiv
p_{\parallel} /p_F$. For the lower sign appropriate to
$u_{\uparrow}-v_{\uparrow}$, the solutions for $x$ and hence
$p_{\|}$ is given by $p_F [+{i}\frac{\Delta_0}{2 E_F}\pm
\sqrt{\frac{\tilde \mu}{E_F}}]$ (assuming a weak-coupling
superconductor and thus $\Delta_0 \ll E_F$) with positive imaginary
parts (if $\tilde \mu
>0$).  Hence $u_{\uparrow}-v_{\uparrow}$ has a pair of decaying
solutions $\chi_1(\rho)$ and $\chi_2(\rho)$. For the upper sign
appropriate to $u_{\uparrow} + v_{\uparrow}$, we only have two
growing solutions that must be rejected and so
$u_{\uparrow}({\rho})+v_{\uparrow}({\rho})=0$ must hold. Now both
solutions $\chi_1(\rho)$ and $\chi_2(\rho)$ in general contain a
divergence of $\ln\rho$ as $\rho\rightarrow{0}$. Nevertheless, a
finite solution as $\rho\rightarrow0$  can be found by an
appropriate linear combination so that we can write the zero-energy
state as the first line in (\ref{solution1}),
$R_1({\rho})=a\chi_1(\rho)+b\chi_2(\rho)$ where the coefficients are
determined to cancel $\ln\rho$ near the origin. (For $|p_z| > p_F$,
$\tilde \mu < 0$, then both $u_{\uparrow} \pm v_{\uparrow}$ have a
single solution which decays at infinity and thus no zero-energy
state is allowed, {\it c.f.}, e.g., \cite{RG,Tewari}).

The state associated with $\alpha^{\dag}_2$ can be obtained in a
similar manner. The factorization is
$(u_{\downarrow}(\vec{r}),u_{\downarrow}(\vec{r}))
=e^{ip_zz}(e^{i\phi}\tilde{u}_{\downarrow}({\rho}),e^{-i\phi}\tilde{v}_{\downarrow}({\rho}))$.
The equations are also decoupled for
$\tilde{u}_{\downarrow}\pm\tilde{v}_{\downarrow}$ as in previous
case, but the kinetic energy contains an additional term
$-1/\rho^2$, which causes a divergence of $1/\rho$ for the solution
near the origin. Analogous procedure shows that there are two
decaying solutions $\eta_1(\rho)$ and $\eta_2(\rho)$ for
$\tilde{u}_{\downarrow}-\tilde{v}_{\downarrow}$, but two
exponentially increasing solutions for
$\tilde{u}_{\downarrow}+\tilde{v}_{\downarrow}$. Therefore we have
the solution as
$(\hat{u}_2(\vec{r}),\hat{v}_2(\vec{r}))=e^{ip_zz}(0,e^{i\phi},0,-e^{-i\phi})R_2({\rho})$,
where $R_2=c\eta_1+d\eta_2$ is a suitable linear combination to
cancel the divergence of $1/\rho$. Now we obtain the eigenfunctions
associated with the two zero-energy excitations.
We note the relation

\be
    \hat{u}_2=e^{i\phi}\sigma_1\hat{u}_1\kappa(\rho)\:,\
    \hat{v}_2=e^{-i\phi}\sigma_1\hat{v}_1\kappa(\rho),\:\label{identity1}
\ee where $\kappa(\rho)\equiv{R}_2(\rho)/R_1(\rho)$, which will be
useful later.

With the solutions (\ref{solution1}) obtained for a pure $p$-wave
order parameter (\ref{p_order}), we are going to consider the
effects of lacking inversion symmetry on (\ref{solution1}). For
simplicity of presentation, we shall consider separately the
Rashba interaction and an admixture of singlet order parameter but
we only state the general conclusion at the end. First, we shall
show via perturbation theory that a small Rashba interaction or a
small s-wave order parameter would not destroy the zero energy
states. Then we shall consider general magnitude of these two
interactions.

Now we add a Rashba  spin-orbital interaction
$h=(-\alpha\hat{\rm{z}}\times{\bf{p}}\cdot\vec{\sigma})$ to the the
kinetic energy parts $H_0$ in (\ref{BdG}). In cylindrical
coordinates, $h$ and its counterpart $-h^{*}$ associated with the
hole sector can be written as

\bea
    \pm\alpha\left(\begin{array}{cccc}&0&
    e^{\mp{i}\phi}(\partial_{\rho}\mp\frac{i}{\rho}\partial_{\phi})\\
    &-e^{\pm{i}\phi}(\partial_{\rho}\pm\frac{i}{\rho}\partial_{\phi})&0\\
    \end{array}\right)\:,\label{spin_orbit}
\eea where the upper(lower) sign is for the electron(hole) sectors,
respectively. The expectation values of the this spin-orbital
interaction for either of the two states given above are obviously
zero. The matrix element between the two states in (\ref{solution1})
is proportional to the spatial integral of
$\hat{u}^\dag_2h\hat{u}_1-\hat{v}^\dag_2h^{*}\hat{v}_1$, which, with
eq (\ref{identity1}), equals
$\{\hat{u}^\dag_1\sigma_1[e^{-i\phi}{h}+e^{i\phi}(-h^{*})]\hat{u}_1\}$
times some function of $\rho$. Hence the matrix element is zero by
explicit use of (\ref{spin_orbit}). Therefore, within perturbation,
the two states in (\ref{solution1}) are unaffected by the coupling
from the spin-orbital interaction.

Now consider an additional a singlet pairing order parameter. We
therefore add  a term

\be
    \Delta_s(\vec{r})=\left(\begin{array}{cccc}&0&{e}^{i\phi}\\
    &-{e}^{i\phi}&0\end{array}\right)\Delta_s({\rho})\:.\label{s_order}
\ee to the pure triplet one in (\ref{p_order}) in eq (\ref{BdG}).
For mixing which do not break time reversal symmetry far from the
vortex, the ratio $\lim_{\rho \to \infty}
\left(\Delta_s(\rho)/\Delta_p(\rho)\right) \equiv \beta$
 must be real. The expectation value of (\ref{s_order}) in any
 given state in (\ref{solution1}) is again obviously zero, and
 matrix element of
(\ref{s_order}) between the two states in (\ref{solution1}) involves
the spatial integral of
$\hat{u}_2^{\dag}e^{i\phi}(i\sigma_2)\hat{v}_1+\hat{v}_2e^{-i\phi}(-i\sigma_2)\hat{u}_1$
which, by using (\ref{identity1}), is zero. Hence in the small
$\beta$ regime, the two zero-energy states remain.

The previous two paragraphs demonstrate that the two states in
(\ref{solution1}) remain according to perturbation theory. Now we
consider the more general case.
 We shall show that
there are two zero-energy states with general form

\be
    e^{ip_zz}\left[u_{\uparrow}({\rho}),e^{i\phi}\tilde{u}_{\downarrow}({\rho}),
v_{\uparrow}({\rho}),e^{-i\phi}\tilde{v}_{\downarrow}({\rho})\right]^T\:,\label{solution2}
\ee with $u_{\uparrow}=-v_{\uparrow}$ and
$\tilde{u}_{\downarrow}=-\tilde{v}_{\downarrow}$  all finite and
decaying at infinity, which survive the additional interactions.

First consider order parameter mixing. The operator $\Pi$ in BdG
equation is a sum of (\ref{p_order}) and (\ref{s_order}). The
corresponding set of differential equations are:

\bea
    L_0{u_{\uparrow}}-Pv_{\uparrow}-\Delta_s(\rho)\tilde{v}_{\downarrow}=0\:,\nonumber\\
    L_0{v_{\uparrow}}-Pu_{\uparrow}-\Delta_s(\rho)\tilde{u}_{\downarrow}=0\:,\nonumber\\
    L_1{\tilde{u}_{\downarrow}}-P\tilde{v}_{\downarrow}+\Delta_s(\rho){v_{\uparrow}}=0\:,\nonumber\\
    L_1{\tilde{v}_{\downarrow}}-P\tilde{u}_{\downarrow}+\Delta_s(\rho){u_{\uparrow}}=0\:,\label{BdG_1}
\eea where the differential operators
$L_n=\frac{1}{2m}(\frac{d^2}{d{\rho}^2}+\frac{1}{\rho}\frac{d}{d{\rho}}-\frac{n^2}{{\rho}^2})
+\tilde \mu$ and
$P=\frac{\Delta({\rho})}{p_F}(\frac{d}{d{\rho}}+\frac{1}{2{\rho}})$.
Denoting $w^{\pm}_{\uparrow}=u_{\uparrow}\pm{v_{\uparrow}}$ and
$w^{\pm}_{\downarrow}=\tilde u_{\downarrow}\pm{\tilde
v_{\downarrow}}$, the above equations for the $w^{+}$'s are
decoupled with the $w^{-}$'s. Writing
$z^{+}_{\pm}=w^{+}_{\uparrow}{\pm}\ iw^{+}_{\downarrow}$ and
$z^{-}_{\pm}=w^{-}_{\uparrow}{\pm}\ iw^{-}_{\downarrow}$, we arrive
at

\bea
    (L_0+P-\frac{1}{4m{\rho}^2}\mp{i\Delta_s({\rho})})z^{-}_{\pm}+\frac{1}{4m{\rho}^2}z^{-}_{\mp}=0\:,\\
    (L_0-P-\frac{1}{4m{\rho}^2}\pm{i\Delta_s({\rho})})z^{+}_{\pm}+\frac{1}{4m{\rho}^2}z^{+}_{\mp}=0\:.
\eea Though in general $z^-_+$ ($z^+_+$) couples with $z^-_-$
($z^+_-$),  we note that at infinity the above set of equations for
each of $z^{+}_{\pm}$ and $z^{-}_{\pm}$ couples to itself only. Then
for a given $p_z=p_F\cos\theta$ associated with the excitation, we
can write $\tilde \mu=E_F\sin^2\theta>0$.
 $z^{-}_{\pm}$ satisfies, asymptotically for large $\rho$,

\be
    \left[\frac{1}{2m}\frac{d^2}{d{\rho}^2}+\frac{\Delta_0}{p_F}
    \frac{d}{d\rho}+E_F\sin^2\theta\mp{i}\beta\Delta_0\right]z^{-}_{\pm}=0\:,\label{asymp}
\ee where one recalls that $\Delta_s(\rho\rightarrow\infty) = \beta
\Delta_0$. Note that a similar equation as (\ref{asymp}) except the
positive coefficient in front of $d/d\rho$ is for the $z^{+}$'s.
Again, take $e^{ip_{\|}\rho}$ as asymptotic solutions. In the
weak-coupling limit, $p_{\|}$ for $z^{-}$ are given by
$p_F[i\frac{\Delta_0}{2
E_F}\pm\sqrt{\sin^2\theta\pm{i}\beta\frac{\Delta_0}{E_F}}]$.
 The imaginary parts are therefore, for small $\Delta_0/E_F$,
$\frac{p_F}{2}\frac{\Delta_0}{E_F}(1\pm\frac{\beta}{\sin\theta})$.
For a given $p_z$ and $\theta$, there are four roots associated with
$z^{-}_{\pm}$ with positive imaginary parts when
$\frac{|\beta|}{\sin\theta}<1$, which leads to four independent
decaying solutions for $z^{-}_{\pm}$. The same arguments show that
the solutions for $z^{+}_{\pm}$ are exponentially increasing as
$\rho \to \infty$,  and hence we must choose
$w^{+}_{\uparrow}=w^{+}_{\downarrow}=0$, that is,
$u_{\uparrow}=-v_{\uparrow}$ and
$\tilde{u}_{\downarrow}=-\tilde{v}_{\downarrow}$ as in
(\ref{solution2}). Therefore a general solution for the zero-energy
state for Eq.\ (\ref{BdG_1}) can be represented by the linear
combination of the 4 decaying solutions,

\be
    \left(\begin{array}{cccc}u_{\uparrow}(\rho)\\
    \tilde{u}_{\downarrow}(\rho)\\
    v_{\uparrow}(\rho)\\
    \tilde{v}_{\downarrow}(\rho)
    \end{array}\right)=
    \sum_{i=1}^4c_i
    \left(\begin{array}{cccc}1\\y_i(\rho)\\-1\\-y_i(\rho)
    \end{array}\right)f_i(\rho)\:,
\ee where the $f$'s decays towards zero at infinity, and so are
the $y_if_i$. From eq (\ref{BdG_1}), the divergences near the
origin are of the form $\ln\rho$ for the first row, and $1/{\rho}$
for the second row. Now we have two equations determining the
$c$'s by which the respective divergence can be removed. In
general, we can have two independent sets of $\{c_i\}$ satisfying
the above, which in turn leads to two independent zero-energy
states within the vortex core. A crucial consequence is drawn from
the above arguments. For the relative pairing strength
$|\beta|>1$, $|\beta/\sin\theta|>1$ so that the zero-energy states
no longer survive, which we conclude that, in addition to $\tilde
\mu=0$, $|\beta_c|=1$ is another critical parameter. On the other
hand, the zero-energy states exist at the core when $|\beta|<1$.
The density of such excitations (number per unit length of the
vortex line) can  be determined from the condition
$|\beta/\sin\theta|<1$. For a spherical Fermi surface, we obtain
$\frac{2p_F}{\pi}\sqrt{1-\beta^2}$. We note that the energy gap
for the bulk excitations are given by $\Delta_0 |{\rm sin} \theta
\pm \beta|$, hence the above critical value of $\theta$ for the
existence of the $E=0$ vortex bound state corresponds to exactly
the existence of a nodal line in one of the branches.  This is
reasonable as this is the value of $p_z$ where two of the decaying
solutions for ${\rm sin} \theta
> |\beta|$ become extended, destroying the possibility of obtaining
the solution eq (\ref{solution2}) which converges both at $\rho \to
\infty$ and $0$.

Next we move on to the case when the spin-orbital interaction
(\ref{spin_orbit}) is included in (\ref{BdG}). With again the
wavefunction in the form of eq (\ref{solution2}), the
zero-energy BdG equation
can be written as

\bea
    L_0u_{\uparrow}-Pv_{\uparrow}-\alpha(\frac{d}{d\rho}+\frac{1}{\rho})\tilde{u}_{\downarrow}=0\:,\nonumber\\
    L_0v_{\uparrow}-Pu_{\uparrow}-\alpha(\frac{d}{d\rho}+\frac{1}{\rho})\tilde{v}_{\downarrow}=0\:,\nonumber\\
    L_1\tilde{u}_{\downarrow}-P\tilde{v}_{\downarrow}+\alpha\frac{d}{d\rho}u_{\uparrow}=0\:,\nonumber\\
    L_1\tilde{v}_{\downarrow}-P\tilde{u}_{\downarrow}+\alpha\frac{d}{d\rho}v_{\uparrow}=0\:.
\eea We can analyze these equations in the same manner as the
previous case. At infinity the BdG equations become decoupled as

\be
    \left[\frac{1}{2m}\frac{d^2}{d{\rho}^2}+(\frac{\Delta_0}{p_F}
    {\pm}i\alpha)\frac{d}{d\rho}+E_F\sin^2\theta\right]z^{-}_{\pm}=0\:,
\ee where $\cos\theta=p_z/p_F$, and note there is a similar equation
for $z^{+}_{\pm}$ except the overall positive coefficient associated
with $d/d\rho$. $p_{\|}$'s associated with the asymptotic solution
$e^{ip_{\|}\rho}$ have imaginary part
$\frac{\Delta_0}{2E_F}(1\pm\frac{\alpha/v_F}{\sqrt{(\alpha/v_F)^2+\sin^2\theta}})$,
which is positive for all $\theta$. (Here $v_F \equiv p_F/m$). On
the contrary, the corresponding $p_{\|}$ for $z^{+}_{\pm}$ have only
negative imaginary parts. Applying the same arguments as before, the
zero-energy bound states survive under any magnitude of the
spin-orbital interaction. One can understand this result by the fact
that the size of the Fermi surfaces at $p_z$ for the two branches
are given by  $p_{F \pm} \equiv \left[ ( 2 m \tilde \mu) + (m^2
\alpha^2) \right]^{1/2} \pm m \alpha $
 $ = p_F \left[ \sqrt{(\frac{\alpha}{v_F})^2+\sin^2\theta} \pm \frac{\alpha}{v_F} \right]$
which remain finite for arbitrary large value of $\alpha$.

The above analysis can be generalized to the case where both
$\alpha$ and $\beta$ are finite. We find that a pair of $E=0$
states exist if one has both
$\sqrt{(\frac{\alpha}{v_F})^2+\sin^2\theta} + \frac{\alpha}{v_F} +
\beta > 0$ and $\sqrt{(\frac{\alpha}{v_F})^2+\sin^2\theta} -
\frac{\alpha}{v_F} - \beta > 0$.  We note that in the helicity
basis (spin quantization axis along $(\hat z \times {\bf p})$),
the order parameter on the $\pm$ branches of the Fermi surfaces
are given respectively by $\Delta_p \frac{p_{F \pm}}{p_F} \pm
 \Delta_s =
 \Delta_p \left[\sqrt{(\frac{\alpha}{v_F})^2+\sin^2\theta} \pm \frac{\alpha}{v_F}
 \pm \beta \right]$,
hence the existence or absence of the $E=0$ bound states is
determined by the relative sign of the order parameter on these
two Fermi surfaces.

The solutions in (\ref{solution2}) taking the lack of inversion
into account are as robust as that of (\ref{solution1}) in p-wave
superfluids. It is evident that the local charge density is zero
since the solutions have equal magnitudes for electron and hole
excitations of the same spin projection. Hence the states are not
susceptible to nonmagnetic impurities. Similarly, the states are
not affected by the Zeeman magnetic field along $\hat z$ again due
to the particle-hole symmetry.  Furthermore, they are also not
altered by exchange or Zeeman fields in the in-plane directions,
for the azimuthal dependence in each of (\ref{solution2}) leads to
zero matrix elements of the local spin density operator among the
states\cite{footnote}.

With the two general independent solutions of the form in
(\ref{solution2}),  the corresponding creation operators are not
necessarily self-hermitian. Here we show that a set of two
independent Majorana fermions can be built from them. We first
demonstrate this for $p_z = 0$. From the two linearly independent
solutions, one first construct two orthonormal wavevectors $(\hat
u, \hat v)$ and thus two corresponding operators $\alpha^{\dag}_1$
and $\alpha^{\dag}_2$. From orthonormal properties one then have
$\{\alpha^{\dag}_1,\alpha_1\}=\{\alpha^{\dag}_2,\alpha_2\}=1$ and
$\{\alpha^{\dag}_1,\alpha_2\}=0$. Since $[H_{\rm
eff},\alpha^{\dag}_i]=E_i\alpha^{\dag}_i=0$, we also have $[H_{\rm
eff},\alpha_i]=-E_i\alpha_i=0$ for $i=1,2$.  Since we have only
two linearly independent solutions to eq (\ref{BdG}) of zero
energies, $\alpha_{1,2}^{\dag} $ must just be a linear
combinations of $\alpha_{1,2}$.  We denote this in matrix notation
as ${\bf \alpha}^{\dag}=\bf{C \alpha}$ where ${\bf C}$ is a $2
\times 2$ matrix. We shall find the transformation ${\bf
\gamma}^{\dagger} = {\bf W \bf \alpha}^{\dagger}$ such that the
$\gamma$'s are independent Majorana fermion operators, that is,
$\gamma^{\dag}_i=\gamma_i$ for $i=1,2$, and
$\{\gamma_1,\gamma_2^{\dag}\} = 0$. By choosing $\bf{W}$ as
unitary, we can assure the conditions of normalization and
orthogonality for the $\gamma$'s. It remains only to make the
$\gamma$'s self-conjugate.
 From  $\{\alpha_i,
\alpha_j^{\dagger}\} = \delta_{ij}$ where $i,j$ $=1$ or $2$, one can
show that $\bf{C}$ is unitary. Furthermore,
$\alpha^{\dag}=\bf{CC^{*}}\alpha^{\dag}$ and thus
${\bf{C}}^{-1}=\bf{C^{*}}$. Since ${\bf C}$ is unitary, $\bf{C}$ is
also symmetric. Hence we can write
${\bf{C}}=e^{i\lambda}e^{i\omega\hat{n}\cdot\vec{\sigma}}$, where
$\lambda$ is real and $\hat{n}$ is perpendicular to $\hat{\rm{y}}$.
 We thus have
${\bf \gamma ^{\dag}}={\bf{WC}}\alpha={\bf{WCW}}^T{\bf \gamma}$
which would equal $ {\bf \gamma}$ if we choose
${\bf{W}}=e^{-i\frac{\omega}{2}\hat{n}\cdot\vec{\sigma}}e^{-i\lambda/2}$.
Thus the operators $\gamma^{\dag}_1,\gamma^{\dag}_2$ constitutes a
set of two independent Majorana fermions. The constructions of
Majorana fermions for finite $p_z$'s can proceed in a similar manner
if we replace the $e^{i p_z z}$ factors in the wavefunctions by
${\rm cos} (p_z z)$ and ${\rm sin} (p_z z)$.

As demonstrated already in, e.g., \cite{RG,Ivanov}, the dimension
of Hilbert space of a Majorana fermion is $\sqrt{2}$, that is,
each two Majorana fermions combines to form one Fermionic state
with two degrees of freedom (occupied or empty).  Hence, for our
system with $n_v$ vortices per unit area, we have a residual
entropy density $n_v \frac{p_F}{\pi} \sqrt{1 - \beta^2 -
\frac{2\alpha}{v_F} \beta} \ln 2 $. This ground state degeneracy
is lifted only by the finite overlap between the vortices
\cite{Radzihovsky}, the resulting energies are thus exponentially
small in the vortex spacings. The existence of this residual
entropy can be used to demonstrate the existence of $E=0$ vortex
bound states, as well as be a measure of the mixing of the two
superconducting order parameters.

In conclusion, we have considered the vortex bound states in a
non-centrosymmetric superconductors, in particular for an order
parameter appropriate to CePt$_3$Si.  We demonstrated that the
zero energy states exist only for certain range of $p_z$ values
depending on the magnitudes of the singlet versus the triplet
order parameters.

While preparing this manuscript, we become aware of
\cite{Fujimoto08} which however discusses a very different aspect
of vortex bound states of non-centrosymmetric superconductors.

This research was supported by the National Science Council of
Taiwan under grant number NSC95-2112-M001-054-MY3.

\end{document}